\documentclass[%
 reprint,
 amsmath,amssymb,
 aps,
]{revtex4-2}

\usepackage{graphicx}
\usepackage{dcolumn}
\usepackage{bm}
\usepackage{qcircuit}
\newcommand{\bra}[1]{\langle #1 |}
\newcommand{\ket}[1]{| #1 \rangle}
\newcommand{\ham}{\mathcal{H}}
\newcommand{\avg}[1]{\langle #1 \rangle}
\newcommand{\eo}{\mathcal{E}_{0}}
\newcommand{\ef}{\mathcal{E}_{f}}
\newcommand{\nrg}{\mathcal{E}}

\newcommand{\sz}{\sigma^{z}}
\newcommand{\sx}{\sigma^{x}}

\newcommand{\chisq}{\chi^{2}/\text{d.o.f.}}

\begin{document}


\title{Metropolis-style random sampling of quantum gates for the estimation of low-energy observables}
\author{Judah F. Unmuth-Yockey}
\email{jfunmuthyockey@gmail.com}
\affiliation{Theory division,
Fermi national accelerator laboratory,
Batavia, IL 60510-5011, USA}

\date{\today}

\begin{abstract}
We propose a quantum algorithm to compute low-energy expectation values of a quantum Hamiltonian by sampling a partition function associated with the average energy of that Hamiltonian.  For any given quantum circuit-Hamiltonian pair, there is an associated average energy.  The sampling is done through an accept/reject Metropolis-style algorithm on the quantum gates of the circuit itself.  Observables calculated under the canonical ensemble from these samples of circuits are extrapolated from higher-energies to the ground state.
\end{abstract}

\maketitle


\section{\label{sec:intro}Introduction}
In many areas of physics, we are often interested in calculating observables when the physical system is in, or is around, its lowest energy state.
One particularly difficult problem of this nature is the calculation of expectation values of observables in the ground state of a quantum system.
Finding the eigenvector with the smallest eigenvalue for a many-body quantum Hamiltonian using a classical computer rapidly becomes expensive as the number of bodies becomes large~\footnote{This is due to the exponential growth of the Hilbert space with system size.  Symmetries and conservation laws, along with sparse storage, can alleviate this cost.}.
Quantum computing potentially alleviates some of this expense.
Algorithms on quantum computers to compute the ground state of a Hamiltonian have provided a vast speed up---a quadratic improvement---over classical algorithms~\cite{grover1996,poulin2009}.
Often, however, we are not necessarily interested in the actual eigenvector itself; it is rather a means to an end: to calculate an  average of some observable.

A powerful tool to calculate gross properties---like averages---nonperturbatively comes from Markov chain Monte Carlo sampling.  
In the case of a classical statistical mechanics system, one can sample the configurations of a partition function to reconstruct ensemble averages under the canonical ensemble.
This is what is done in lattice gauge theory, for example, where a system of fundamental particles are simulated on a Euclidean lattice such that gauge invariance is maintained exactly~\cite{montvay_munster_1994}.  
Average properties of the system can be computed and, when possible, Wick-rotated back into Minkowski space-time.  
In this context, it is important to take the continuum (and infinite-volume) limit.  
This is done by taking the gauge-coupling, $g$, to zero, or $\beta_{\text{gauge}} \sim 1/g^2 \rightarrow \infty$.

However, in practice one never performs simulations with $\beta_{\text{gauge}}$ infinitely large, and instead moderate values of $\beta_{\text{gauge}}$ are chosen, and the infinite-$\beta_{\text{gauge}}$ limit is done by extrapolation.
In this way, continuum-limit results are found from finite-lattice-spacing calculations.
Here, we propose to perform a similar extrapolation when what is desired are expectation values of observables in the ground state.
We propose to perform systematic calculations of observables at larger energies, and extrapolate to their value at low energy by fitting.
This is not done by a variational minimization routine~\cite{Peruzzo2014,Cerezo2021,Yeter-Aydeniz2021,tilly2021variational,zhang2021simulating,li2021benchmarking}, where the ground state is sought out by minimizing a cost function of expectation values; though, a procedure similar in spirit to what is proposed here is described in Ref.~\cite{cao2021energy}.   The ground state is not sought out using quantum imaginary time evolution~\cite{Motta2020,Yeter_Aydeniz_2021,Yeter-Aydeniz2021} which involves a classical minimization routine based on state tomography to construct unitary circuits to implement non-unitary operators.  We do not use a strictly ``gate-based'' method, where the ground state of a quantum Hamiltonian is calculated with some finite probability~\cite{ge2019,Lin2020nearoptimalground,Wang2017,LEE2020135536,gustafson2020}.  The method proposed here is also not a quantum Metropolis algorithm, where a thermal quantum state is sampled~\cite{Temme2011,Yung754,moussa2019measurementbased}, or a state is adiabatically annealed~\cite{Kirkpatrick1983,kadowaki1998,farhi2001,Hauke_2020} to its ground state.
Instead, by using samples of averages of observables calculated away from the ground state, we extrapolate those observables to their lowest-energy values with some error, avoiding the ground state entirely.

The remainder of the article is as follows:  In Sec.~\ref{sec:random-circuit-sampling} we describe the algorithm in detail, and discuss its computational cost.  In Sec.~\ref{sec:ising} we use the algorithm in an example, and compute the ground-state energy eigenvalue and magnetization for the transverse-field quantum Ising model for a small system using a classical computer.  Finally in Sec.~\ref{sec:conclusions} we provide possible future directions and work, as well as conclusions.

\section{Random circuit sampling}
\label{sec:random-circuit-sampling}
The goal of the algorithm is to construct an ensemble of \emph{circuits} whose average energies fluctuate about some
equilibrium value controlled by an external parameter set by the user.  This is done non-perturbatively by the random sampling of quantum gates.  By controlling the target equilibrium average-energy value, observables of interest can be calculated from the averages of the circuits, and extrapolated to the low-energy sector.
This is possible because of the functional form of the Boltzmann weight and the partition function, where smaller ``temperatures'' favor lower energy configurations.  In addition, because of the variational principle, energy expectation values are upper-bounds on the ground state energy, so that when extrapolating to lower energies, this bound is satisfied.
In this way, any state preparation is unnecessary, and instead one extrapolates the physics of interest from higher energy expectation values, to their low-energy values.  We now describe the algorithm.

We begin with an initial state $\ket{\psi_{0}}$ for $N_{q}$ qubits that is easy to prepare, such as the all-`plus' state, or the all-zeros state.
We assume the circuit---represented by the unitary $U$---that acts on this state can be expressed in $N_{t}$ layers, that is, $U = U^{(N_{t})} U^{(N_{t}-1)} \cdots U^{(2)} U^{(1)}$. Each layer has three sub-layers.  There is a sub-layer of one-qubit gates, and two sub-layers of two-qubit gates, where each two-qubit sub-layer acts on distinct pairs of qubits.
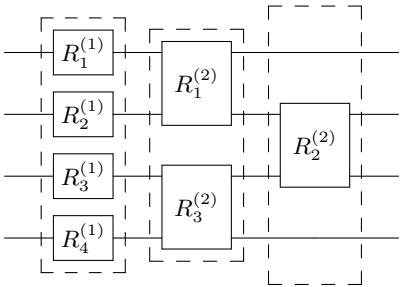
\begin{figure}[ht]
\mbox{
\Qcircuit @C=2em @R=0.7em {
& \gate{R^{(1)}_{1}} & \multigate{1}{R^{(2)}_{1}} & \qw & \qw \\
& \gate{R^{(1)}_{2}} & \ghost{R^{(2)}_{1}} & \multigate{1}{R^{(2)}_{2}} & \qw \\
& \gate{R^{(1)}_{3}} & \multigate{1}{R^{(2)}_{3}} & \ghost{R^{(2)}_{2}} & \qw \\
& \gate{R^{(1)}_{4}} & \ghost{R^{(2)}_{3}} & \qw & \qw
\gategroup{1}{2}{4}{2}{1em}{--}
\gategroup{1}{3}{4}{3}{1em}{--}
\gategroup{1}{4}{4}{4}{3.8em}{--}
}
}
\caption{A single layer with four qubits, comprised of three sub-layers which are boxed with dashed lines.  One sub-layer only has single-qubit rotations indicated by $R^{(1)}$, while the next two sub-layers have two-qubit rotations, indicated by $R^{(2)}$.}
\label{fig:sub-layers}
\end{figure}
Figure~\ref{fig:sub-layers} shows an illustration of the layer structure.  The one- and two-qubit gates are initialized randomly, or \emph{e.g.} as identity matrices, from the groups $U(2)$ and $U(4)$, respectively.  For the description here we also assume that each of the $N_{t}$ layers is identical. \emph{I.e.} if we know the gates in a single layer, we know them in all layers.  This is simply the statement that the global unitary which is acting on $\ket{\psi_{0}}$ can be decomposed via Trotterization, and does not have any ``time dependence.''  At this point, a model must be chosen whose physics is the physics of interest.  We specify this model with a Hamiltonian, $\ham$, which is Hermitian.  

We now proceed using a Metropolis-style update~\cite{metropolis} method for the quantum gates in the circuit.  The Metropolis algorithm consists of proposing a new circuit configuration, and accepting that new circuit configuration based on a probability given by the Boltzmann weight associated with the change in energy between the proposed, and current configurations~\cite{berg2004}.  With the initial state, and the initial circuit, we evolve the initial state using the circuit many times to calculate the average energy associated with that circuit, $\eo \equiv \bra{\psi_{0}} U^{\dagger} \ham U \ket{\psi_{0}} = \avg{U^{\dagger} \ham U}$~\footnote{The circuit must be run as many times as is necessary to achieve the desired degree of accuracy on the energy expectation value.  The scaling of this this accuracy is described at the end of the section.}.  We then choose a gate at random in a layer (it can be a fixed layer, \emph{e.g.} the first layer, since we enforce that all layers be the same), and for that gate we propose a new unitary matrix sampled uniformly from the Haar measure~\cite{Hurwitz1897,Haar1933}~\footnote{One could also propose a new unitary somewhat `close' to the current unitary to improve acceptance.}.  It may be preferable to bias how one chooses the gates to update.  This would need to be investigated based on other factors \emph{e.g.} auto-corelations, acceptance, and circuit ansatz.  We then evolve the initial state with the proposed modified circuit---expressed as $U'$---many times to compute the new expectation value, $\ef \equiv \avg{ {U'}^{\dagger} \ham U'}$.  Using the current and proposed average energies we accept the new circuit with a probability, $p$, given by
\begin{align}
\label{eq:metropolis-prob}
    p =
    \begin{cases}
    1 & \text{if $\ef \leq \eo$} \\
    e^{-\beta (\ef - \eo)} & \text{otherwise.}
    \end{cases}
\end{align}
In Eq.~\eqref{eq:metropolis-prob} the parameter $\beta$ is chosen by the user, and controls the equilibrium average energy value.  Larger values of $\beta$ push the target average energy lower.  If the new circuit is chosen, the circuit is kept and the process repeats; otherwise, the proposed circuit is ignored, the altered gate is reverted back, a new gate is chosen at random and the process repeats.  Sampling performed in this manner attempts to approximately compute averages with respect to the partition function,
\begin{align}
    Z = \int dV e^{-\beta \avg{V^{\dagger} \ham V}} = 
    \int dV e^{-\beta \nrg(V)}
\end{align}
where $dV$ is the Haar measure for the group $U(2^{N_{q}})$.  Because of the invariance of the Haar measure the partition function is independent of $\ket{\psi_{0}}$.

For a fixed value of $\beta$, at equilibrium, the circuit composition fluctuates such that the average energy is distributed about some equilibrium value, $\overline{\nrg}$.  Using this equilibrium value as a reference, one can choose increasingly larger values of $\beta$, and extrapolate observables to their low-energy values.

The cost of the algorithm is as follows: Since each layer is identical, we need only remember the contents of a single layer, and a proposed gate, which is trivial on a classical computer; however, for $N_{q}$ qubits, and $N_{t}$ layers, there are $N_{q}N_{t}$ arbitrary one-qubit gates in the circuit, and $(N_{q}-1)N_{t}$ arbitrary two-qubit gates in the circuit.    If we define a single ``sweep'' as $N_{q} + (N_{q}-1)$ proposals, and the number of sweeps, $N_{s}$ is chosen by the user, the cost of a sweep---the execution time---scales linearly in the number of qubits.  

Aside from the circuit cost associated with executing arbitrary single- and two-qubit gates, the major overhead comes from the number of measurements. The error, $\epsilon$, on the expectation value of the energy associated with a circuit converges like $\sim 1/\sqrt{N_{\text{meas}}}$ where $N_{\text{meas}}$ is the number of uncorrelated measurements done to compute an expectation value.  Let $\mathcal{N}$ be the number of uncorrelated expectation values recorded.  Then the error, $\lambda$, on the average of expectation values converges similarly like $\sim 1/\sqrt{\mathcal{N}}$.  The final error on the average expectation value scales like $\sim \sqrt{\epsilon^{2} + \lambda^2}$.

\section{The Ising model as an example}
\label{sec:ising}
In this section we attempt to elucidate the procedure described in the previous section with an example: the Ising model.
The one-dimensional, quantum, transverse-field Ising model Hamiltonian with open boundary conditions is given by
\begin{align}
\label{eq:ising-ham}
    \ham_{\text{Ising}} = - \sum_{i=1}^{N_{q}-1} \sz_{i} \sz_{i+1}
    - h_{x} \sum_{i=1}^{N_{q}} \sx_{i},
\end{align}
where $h_{x}$ is the magnitude of the transverse field.  This Hamiltonian will serve as the $\ham$ used in the algorithm.  To test the basic principles behind the algorithm we attempt to classically  compute the ground-state energy eigenvalue, and the expectation value of the magnetization, $M \equiv \sum_{i=1}^{N_{q}} \sx_{i}$, in \emph{the ground state} for a given value of $h_{x}$.  We compare calculations done with the algorithm with exact results calculated using exact diagonalization.

We consider the case of $N_{q} = 4$ and $N_{t} = 6$ (although other layer depths were tried and little dependence was found for this case), and calculate observables at values of the transverse field $h_{x} = 1/4$, and $h_{x} = 3/2$, \emph{i.e.} in the ordered, and disordered phases, respectively.  The circuit, represented by a unitary $U$ in the previous section, is comprised of six layers, and each layer consists of the three sub-layers mentioned before.
We store a single layer as a list with three lists inside of it.  Each list contains the current one-qubit, or two-qubit gates.  A single gate is randomly chosen, and a new gate is proposed.  Then the proposed circuit is made by multiplying the list with itself $N_{t}$ times, the initial state is evolved with that circuit, and the energy expectation value is computed.  

For the classical simulation done here, this energy expectation value is computed exactly.  In practice one would need to evaluate the same circuit many times to compute an accurate value for the energy expectation value.  While this must be done when using a quantum computer, here for convenience we simply calculate the expectation value exactly from the current circuit acting on the initial state.  This new energy expectation value is compared with the current energy expectation value, and the accept/reject procedure is carried out.  The process then repeats.  Notice that in contrast to the classical Ising-spin Metropolis, where a spin-flip is the most dramatic change that can take place; here, the new unitary gate may result is a very different overall unitary evolution.

\begin{figure}
    \centering
    \includegraphics[width=8.6cm]{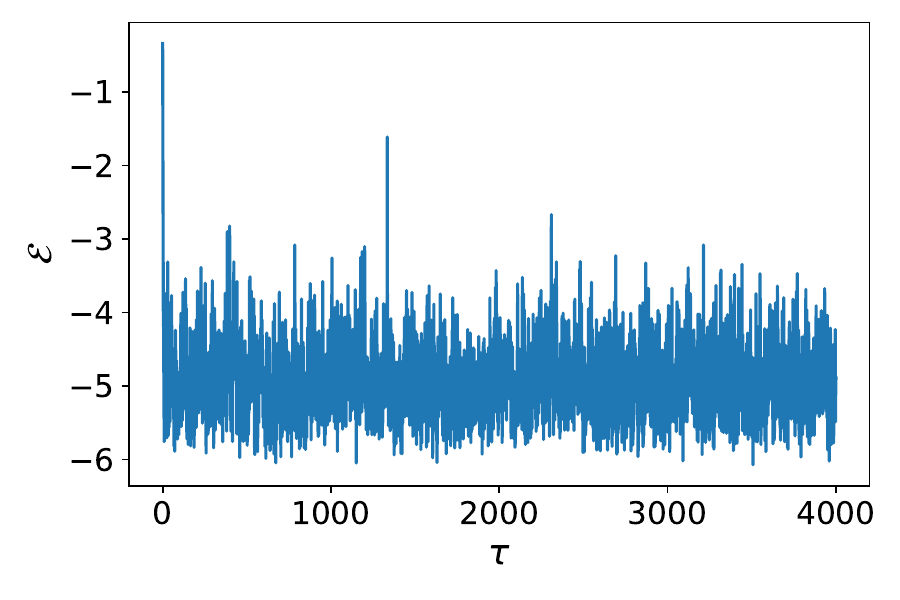}
    \caption{Example time-series data for the expectation value of the energy for the first 4000 measurements.}
    \label{fig:ets}
\end{figure}

We typically perform approximately $1\times10^{6}$ sweeps and record measurements every 10 sweeps to help with auto-correlation.  An example of time series data for the expectation value of the energy can be seen in Fig.~\ref{fig:ets}.
After removing early measurements during equilibration, we bin the data and calculate the error using single-elimination jackknife resampling~\cite{quenouille1956,berg2004} to assess the auto-correlation.  We choose a bin-size when the error saturates, and perform the remainder of the analysis under the jackknife.  

\begin{figure}
    \centering
    \includegraphics[width=8.6cm]{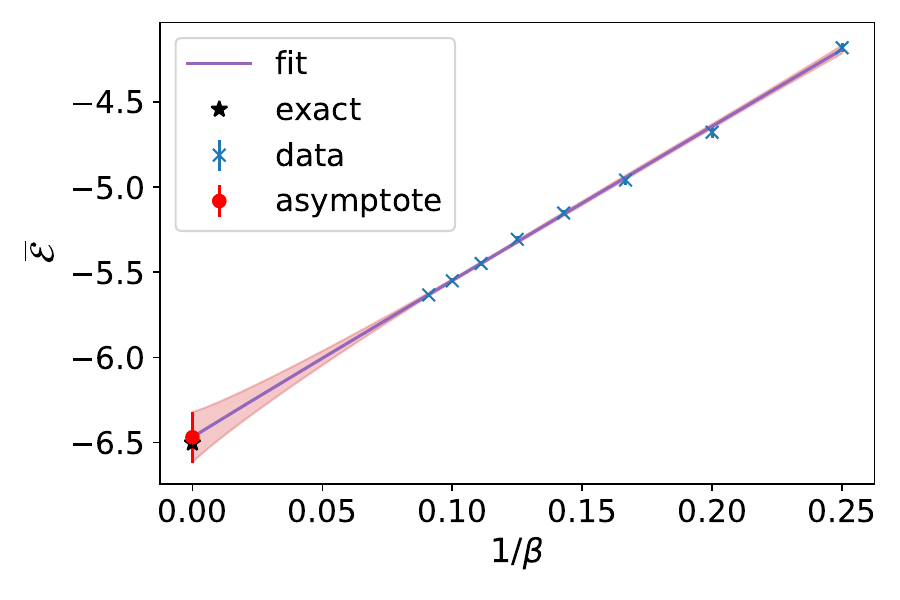}
    \caption{The average energy as a function of $\beta$.  The ground state energy value for the Ising model with $h_{x} = 3/2$ is shown as a black star.  A fit is plotted with its error-band along with the asymptotic value of the fit with error bar.  The largest contribution to the error on the asymptote is statistical.}
    \label{fig:avg-e-1p5}
\end{figure}

\begin{figure}
    \centering
    \includegraphics[width=8.6cm]{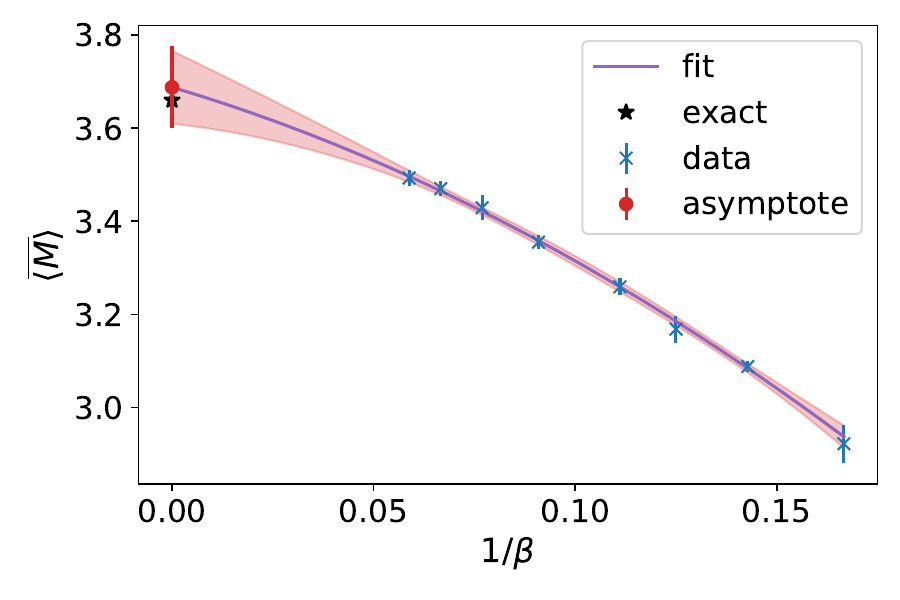}
    \caption{The average of the expectation value of the magnetization for $h_{x} = 3/2$.  Here a fit line is drawn with its error-band, as well as the exact expectation value for the magnetization in the ground state represented as a black star.  The final asymptote to the fit is also drawn with an error bar.  The largest contribution to the error on the asymptote is statistical.}
    \label{fig:avg-m-1p5}
\end{figure}

From the measurements of $\nrg$, and $\avg{M}$ we compute their average values as a function of $\beta$. Measurements of the energy---really the average of the expectation values of the energy---for the case of $h_{x} = 3/2$ can be seen in Fig.~\ref{fig:avg-e-1p5}.
For sufficiently large $\beta$, we find the data approximately obeys a form
\begin{align}
\label{eq:oneoverbeta}
    \overline{\nrg} = \frac{A}{\beta} + B
\end{align}
where $A$ and $B$ are free parameters.  We also allow a floating power-law,
\begin{align}
\label{eq:freepower}
    \overline{\nrg} = \frac{A}{\beta^{C}} + B
\end{align}
with $C$ an additional free parameter, and use the $B$ from this fit to include a systematic error from the fit choice with Eq.~\eqref{eq:oneoverbeta}.  We find good agreement between the data with both Eqs.~\eqref{eq:oneoverbeta} and~\eqref{eq:freepower}, as can be seen in Fig.~\ref{fig:avg-e-1p5}.  The asymptotic value of the fit from Eq.~\eqref{eq:oneoverbeta} gives $\overline{\nrg}_{\infty} = -6.45(10)$ with a $\chisq=0.47$, where the error comes from the statistical error, as well as a small systematic error from the choice of fit and the fit range, which are added in quadrature.  This is to be compared with the exact value of $E_{0} = -6.50$.
For the magnetization with $h_{x} = 3/2$ we find slower convergence, but consistent with a power-law in $\beta$.  Figure~\ref{fig:avg-m-1p5} shows the average of the expectation value of the magnetization.
For this observable we found a fit including quadratic corrections of $1/\beta$,
\begin{align}
\label{eq:quad-corrections}
    \overline{\avg{M}} = \frac{D}{\beta} + \frac{E}{\beta^{2}} + F,
\end{align}
to be most suitable, where $D$, $E$, and $F$ are fit parameters; however, we also tried a floating power-law---as in Eq.~\eqref{eq:freepower}---to assess a systematic error.  Both fits were good, and Fig.~\ref{fig:avg-m-1p5} shows the Eq.~\eqref{eq:quad-corrections} fit, along with the asymptotic value and an error bar.  The choice of fit and fit range contributed a small systematic error to the total asymptote error.  For the asymptote we find $\overline{\avg{M}}_{\infty} = 3.69(8)$, which is to be compared with the exact answer, $\avg{M}_{\text{exact}} = 3.66$.  The fit has a $\chisq = 0.17$.

\begin{figure}
    \centering
    \includegraphics[width=8.6cm]{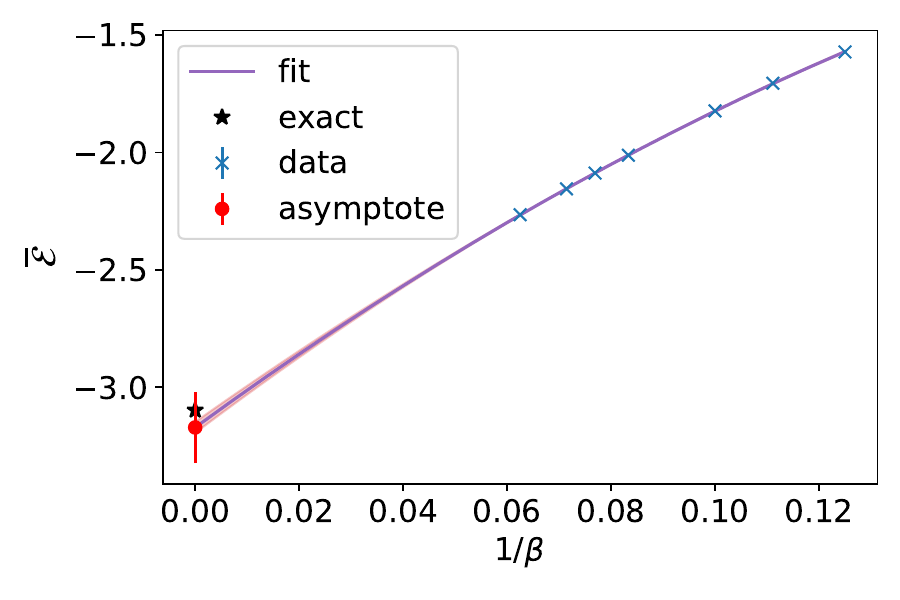}
    \caption{The average energy as a function of $1/\beta$.  The ground state energy value for the Ising model with $h_{x} = 1/4$ is shown as a black star.  A fit is plotted  with its error-band along with the asymptotic value of the fit with error bar.  The largest contribution to the error of the asymptote are the systematic errors associated with the fit ranges and fit functions.}
    \label{fig:avg-e-0p25}
\end{figure}

\begin{figure}
    \centering
    \includegraphics[width=8.6cm]{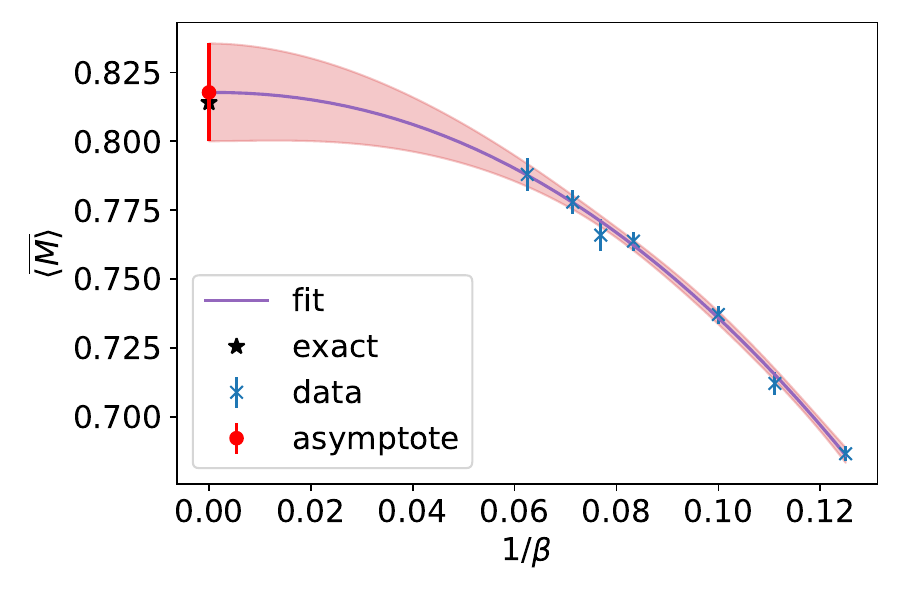}
    \caption{The average of the expectation value of the magnetization for $h_{x} = 1/4$.  Here a fit line is drawn with its error-band, as well as the exact expectation value for the magnetization in the ground state represented as a black star.  The final asymptote to the fit is also drawn with an error bar.  The largest contribution to the error on the asymptote is statistical.}
    \label{fig:avg-m-0p25}
\end{figure}

For $h_{x} = 1/4$, the average energy can be seen in Fig.~\ref{fig:avg-e-0p25}.
We see slight curvature, and so include corrections like $1/\beta^{2}$ into the fit ansatz, in addition to allowing for a single floating power.  The fit with quadratic corrections in $1/\beta$, as well as the exact answer, can be seen in Fig.~\ref{fig:avg-e-0p25}.  We find for the asymptotic value $\overline{\nrg}_{\infty} = -3.17(15)$, which is to be compared with $\nrg_{\text{exact}} = -3.10$.  Here the largest contribution to the error is systematic, with a relatively larger possible range for the asymptote depending on the fit and fit-range used.  The $\chisq = 0.29$.
The magnetization for $h_{x} = 1/4$ can be seen in figure~\ref{fig:avg-m-0p25}.
Due again to the slower convergence, we use the same fits as in the $h_{x} = 3/2$ case, and find we have not really reached the linear region of $1/\beta$ yet.  Nevertheless, the quadratic, and floating-power fits provide a reasonable answer for the asymptote.  The fit and the exact value can be seen in Fig.~\ref{fig:avg-m-0p25}.  We find for the asymptotic value $\overline{\avg{M}}_{\infty} = 0.82(2)$, which is to be compared with the exact value $\avg{M}_{\text{exact}} = 0.814$.  The largest contribution to the error in this case is statistical.  The $\chisq = 0.42$.
We find for this simple system, with only a handful of $\beta$ values, a reasonable extrapolation to the low-energy sector is possible.  

\section{Conclusions and future work}
\label{sec:conclusions}
We have proposed a quantum algorithm to compute expectation values in the low-energy regime of a Hamiltonian.  Rather than use a variational method with a cost-function which is minimized, or attempt to construct an exact circuit which evolves the system to its ground-state directly and use it to calculate observables, we propose to randomly sample circuits, which fluctuate about some target  average energy, and extrapolate measurements to their low-energy values.
The ability to sample around a target average value is a useful feature of this algorithm made possible by fixing the value of $\beta$.  This allows a systematic approach towards lower energies.  The cost for this is degree of control is the introduction of systematic errors in the fitting and extrapolation process, along with run-time overhead due to the nature of probabilistic sampling.

There are a few additional items to note.  The first is how the accuracy of estimates scales.  The most expensive aspect of the algorithm is the slow convergence of averages with the number of samples, which go like one over the square-root of the number of measurements.  Ideally, one could pick a desired error and sample to that extent; however, the situation is more complicated.  Since progressively larger $\beta$ values are needed in the extrapolation, this will lead to a decrease in acceptance during the gate proposal, leading to a necessary increased run-time.  There are many improvements on classical computers for spin and gauge models for the Metropolis algorithm which overcome this issue at ``low-temperature,'' \emph{e.g.} event-driven simulations~\cite{BORTZ197510,berg2004}.  Perhaps similar algorithmic advances could be made here.

Second, as one approaches criticality in a statistical quantum theory, the correlation length diverges.  This typically causes large auto-correlations when one uses the naive Metropolis method, again leading to long run-times.  As before, several algorithms attempt to eliminate this ``critical slowing down,'' \emph{e.g.} cluster flips~\cite{swendsen1987,wolff1989}, worm-like (or world-line) algorithms~\cite{prokofev2001}, both of which are non-local, and hybrid Monte Carlo~\cite{DUANE1987216}.  One could imagine such advances are possible in this case as well.  In addition to the longer run-time, for the ansatz considered here, criticality could push the number of layers to become larger, since the circuit only uses local gates to create long-range correlations.

Finally, the circuit ansatz chosen here is not definitive, and other options could be investigated.  The form proposed here was only done so because of its generality, such that it necessarily can reproduce the full Haar measure sum in the appropriate limit.  However, other gate ansatzes may be more efficient.  This direction deserves more investigation.

The circuit ansatz used here has a physical interpretation in terms of unitary time evolution by an effective Hamiltonian with nearest-neighbor two-spin interactions, and single spins coupled to external fields, since a Hermitian Hamiltonian for the gates can be given in terms of pairs, or single, Pauli operators.  In this sense, during the proposal steps one adjusts the strengths of the nearest neighbor---or on-site---couplings and accepts or rejects them based on the new energy expectation value.  

The possible difficulties associated with the simple accept/reject method could be cured if efficient sampling methods can be devised for the layered-gate ansatz used here.  This ansatz is also a feature which could possibly be improved.  The fact that efficient sampling algorithms have already been devised in the past for spin and gauge systems is promising.  We leave the development of such algorithms as future work.

\begin{acknowledgments}
I would like to thank Hank Lamm, Michael Wagman, Bharath Sambasivam, Jay Hubisz, Erik Gustafson, and Michael Hite for stimulating discussions.  I would also like to thank Hank Lamm, Jack Laiho, Yannick Meurice, and M. Sohaib Alam for reading through drafts of the manuscript.
This manuscript has been authored by Fermi Research Alliance, LLC under Contract No. DE-AC02-07CH11359 with the U.S. Department of Energy, Office of Science, Office of High Energy Physics.
\end{acknowledgments}

%

\end{document}